\documentclass[twocolumn,aps,prb,epsf,graphics,psfig]{revtex4}
\usepackage{graphicx,graphics} 
\usepackage[dvips,bookmarks=false]{hyperref}

\usepackage{graphicx}
\usepackage{dcolumn}

\usepackage{eucal}
\usepackage[dvips]{epsfig}
\begin{document}
\title{Electronic transport through a C$_{60}$ molecular bridge: The role of single and multiple contacts}
\author{Alireza Saffarzadeh$^{1,2,}$}
\altaffiliation{E-mail: a-saffar@tehran.pnu.ac.ir}
\affiliation{$^1$Department of Physics,
Payame Noor University, Nejatollahi Street, 159995-7613 Tehran, Iran \\
$^2$Computational Physical Sciences Laboratory, Department of
Nano-Science, Institute for Research in Fundamental Sciences
(IPM), P.O. Box 19395-5531, Tehran, Iran}
\date{\today}

\begin{abstract}
The effects of different contact geometries, bond dimerization,
and gate voltage on quantum transport through a C$_{60}$ molecule
are studied by the Landauer-B\"{u}ttiker formula based on the
Green's function technique. It is shown that the number of contact
points between the device electrodes and the molecule can play an
important role in the electron conduction. The transmission is due
to the resonant tunneling when the electrodes are contacted to one
carbon atom of the molecule. In the case of multiple contacts, the
interference effects are responsible for the change of the
transmission through the C$_{60}$. The bond dimerization and a
gate voltage shift the molecular levels and by adjusting the
related parameters the electron conduction can be controlled.
\end{abstract}
\maketitle

\section{\bf Introduction}
Molecular electronics using single molecules as active elements is
a promising technological concept with fast growing interest
\cite{Petty,Joach1}. Recent improvements in manipulation of
individual or small numbers of molecules permit us to contact
molecules to metallic electrodes and measure their electronic
transport properties \cite{Reed,Kergueris,Porath,Reichert,Nitzan}.
Electronic transport through single molecules strongly depends on
the nature and quality of the contacts with electrodes. If, e.g.,
a molecule is weakly coupled to the electrodes, the charge at the
molecule becomes strongly localized and the transport takes place
in the regime of Coulomb blockade. In the opposite case (strong
coupling), we can expect to approach the ballistic regime.

The selection of a molecular bridge and the accurate control on
its coupling to the electrodes are basic prerequisites for
designing and manufacturing single molecule electronic devices. In
such a structure, the energies of the molecular orbitals, in
particular, the highest occupied molecular orbital and the lowest
unoccupied molecular orbital (LUMO), are of crucial importance for
the electronic transport through single organic molecules. Among
many types of molecules, the fullerene C$_{60}$ is suitable for
molecular bridge because its LUMO is situated at relatively lower
energies in comparison with the other organic molecules
\cite{Naka}. The electronic structure of isolated C$_{60}$
molecules shows a large gap ($\approx 2$ eV), which indicates that
the molecules should behave like insulators at room temperature
\cite{Dressel}. However, when they are contacted by metallic
electrodes, charge transfer occurs and they become conducting
through the LUMO of the isolated molecules.

During the last decade, the electron conduction through a C$_{60}$
molecule has been extensively investigated both experimentally
\cite{Joach2,Joach3,Por1,Por2,Park,Neel1,Neel2} and theoretically
\cite{Paul,Naka,Mishra,Cunib,Palacios1,Taylor,Palacios2}. Joachim
and co-workers \cite{Joach2,Joach3} studied the conductance
through a C$_{60}$ molecule sandwiched between the Au(110) surface
and a scanning tunneling microscopy (STM) tip at ambient
temperature. The results showed that the current-voltage ($I$-$V$)
characteristics of the molecule are linear at low voltages due to
the absence of molecular orbitals around Fermi energy. Porath and
co-workers \cite{Por1,Por2} deposited isolated C$_{60}$ molecules
onto a gold substrate, covered by a thin insulating layer. In this
way, a double barrier tunnel junction configuration was realized
in which a C$_{60}$ molecule is coupled via two junctions to the
gold substrate and the tip of a STM. The tunneling \textit{I-V}
spectra exhibited a nonvanishing gap in the curves around zero
bias due to the single-electron tunneling effects, such as the
Coulomb blockade and the Coulomb staircase. Park \textit{et al.}
\cite{Park} fabricated single-molecule transistors based on
individual C$_{60}$ molecules connected to gold electrodes. The
device showed strongly suppressed conductance near zero bias
voltage followed by steplike current jumps at higher voltages.
Such a gap could be reduced to zero by adjusting gate voltage.
More recently, N\'{e}el and co-workers \cite{Neel1,Neel2} studied
contacts to a C$_{60}$ molecule on Cu(100) and observed that the
conductance rapidly increases in the transition region from
tunneling to contact, with decreasing tip-molecule distance.

On the other hand, using the tight binding model, it was
theoretically found that the strength of the metal/C$_{60}$
interaction and the geometry of the contact between the tip and
the molecule play an important role in the drastic increase in the
conductance \cite{Paul}. Based on the Green's function method and
the Landauer-B\"{u}ttiker formula, it has been shown that a loop
current emerges in a C$_{60}$ molecule when the electron energy
approaches the energy levels of the molecule. The magnitude of the
such loop currents can be much larger than that of the
source-drain current \cite{Naka}. By incorporating an extra atom
at the center of the fullerene molecule, it is possible to control
the currents in the loops and hence the procedure of transport
\cite{Mishra}. By using density functional theory, the electronic
transport through a C$_{60}$ molecule in between carbon nanotube
leads \cite{Cunib} and Al metallic electrodes has also been
investigated \cite{Palacios1,Taylor,Palacios2}.

The C$_{60}$ molecule consists of 12 pentagons and 20 hexagons.
Due to dimerization, the carbon-carbon bonds in the molecule have
different lengths: $r_1=1.46$ {\AA} for the single bonds (bonds on
the pentagons) and $r_2=1.40$ {\AA} for the double bonds (bonds on
the hexagons that are not shared by a pentagon). We believe that
the effect of bond dimerization on the electron transmission may
considerably affect the $I$-$V$ characteristics, under suitable
conditions. Furthermore, when two electrodes are connected to the
molecule, the number of contact points will be dependent on the
direction of the molecule; thus, single or multiple contacts may
occur. Therefore, the effects of contact geometry in the presence
of bond dimerization and a gate voltage on the electron conduction
through the molecule should be studied. Such features have not
been investigated in the above-mentioned studies. It is the
purpose of this paper to study the role of bond dimerization,
multiple contacts (see Fig. 1), and gate voltage in the coherent
quantum transport through a C$_{60}$ molecule, based on the
tight-binding model and the nonequilibrium Green's function
technique.

The paper is organized as follows. The theoretical model and
formalism are given in section II. In section III, we present the
numerical results of the coherent transport through the C$_{60}$
molecular bridge with different electrode/molecule contacts in the
presence of bias and gate voltages. A brief conclusion is given in
section IV.
\begin{figure}
\centerline{\includegraphics[width=0.5\linewidth]{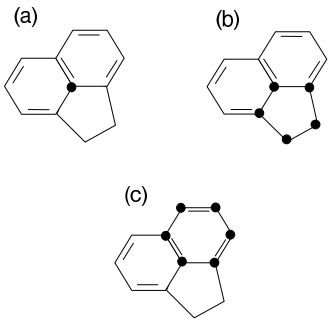}}
\caption{The three different ways of coupling between the C$_{60}$
molecule and the one-dimensional electrodes used in this work.
Black circles show the position and the number of couplings.}
\end{figure}

\section{\bf Model and method}
We consider a system consists of a C$_{60}$ molecule attached to
one-dimensional metallic electrodes. The Hamiltonian of the system
is described within the tight-binding approximation with only one
orbital per atom,
\begin{equation}\label{H}
\hat{H}=\sum_i\epsilon_i\hat{c}_{i}^\dag
\hat{c}_{i}-\sum_{<ij>}t_{ij}\hat{c}_{i}^\dag \hat{c}_{j}\ .
\end{equation}
Here, $\epsilon_i$ is the on-site energy and will be set to zero
except in the gated region (C$_{60}$ molecule) where it is equal
to $V_G$. $\hat{c}_{i\sigma}^\dag$ ($\hat{c}_{i\sigma}$) is the
creation (annihilation) operator for an electron at $i$th site and
$t_{ij}$ is the hopping matrix element between nearest-neighboring
sites $i$ and $j$. The hopping strength in C$_{60}$ molecule
depends on the C-C bond length; thus, we assume different hopping
matrix elements: $t_1$ for the single bonds and $t_2$ for the
double bonds. In the case of bond dimerization, it is reasonable
to use $t_2\simeq(r_1/r_2)^2\,t_1$ \cite{Manou}. The coupling
between the nearest sites in the electrodes is taken to be $t=t_1$
and that between the molecule and the electrodes is taken to be
$t'$. In this study, we assume that the electrons freely propagate
and the only resistance arising from the contacts. This means that
the transport is ballistic \cite{Datta}; therefore, we set
$t'=t/2$ according to Ref. \cite{Naka}.

The Green's function of the C$_{60}$ molecule coupled to the two
metallic electrodes (source and drain) in the presence of the bias
voltage is given as
\begin{equation}\label{G}
\hat{G}_C(\epsilon,V_a)=[\epsilon \hat{1}
-\hat{H}_C-\hat{\Sigma}_L(\epsilon-eV_a/2)
-\hat{\Sigma}_R(\epsilon+eV_a/2)]^{-1}\ ,
\end{equation}
where $\hat{H}_C$ describes the Hamiltonian of the molecule in the
absence of the electrodes and $\hat{\Sigma}_L$ and
$\hat{\Sigma}_R$ describe the self-energy matrices which contain
the information of the electronic structure of the electrodes and
their coupling to the molecule. These can be expressed as
\begin{equation}\label{Self}
\hat{\Sigma}_{L,R}(\epsilon)=\hat{\tau}_{CL,R}\hat{g}_{L,R}(\epsilon)\hat{\tau}_{L,RC}\
,
\end{equation}
where $\hat{g}_{L,R}$ are the surface Green's functions of the
uncoupled leads, i.e., the left and right semi-infinite leads.
$\hat{\tau}$ is a matrix that couples the molecule to the leads
and is determined by the geometry of the molecule-lead bond. Note
that, in the semi-infinite one-dimensional electrodes described by
the single-band tight-binding model, only the first site is
connected to the molecule. As a result, the surface Green's
function for the semi-infinite leads can be written as
${g}_{L,R}=-(1/t)\,e^{ik_{L(R)}a}$, where $k_{L(R)}$ is the wave
vector in the left (right) electrode \cite{Datta}.

\begin{figure}
\centerline{\includegraphics[width=0.8\linewidth]{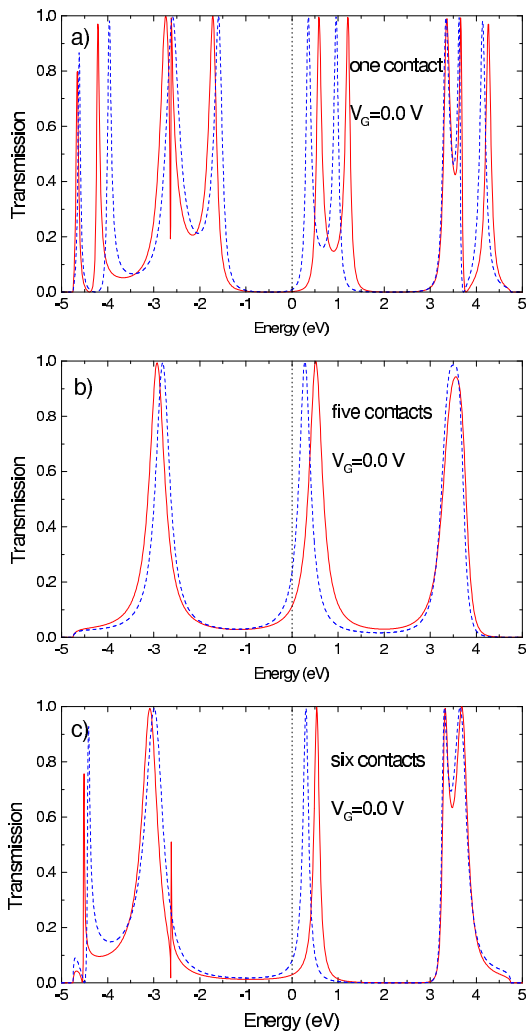}}
\caption{Transmission probability as a function of energy at
$V_G=0.0\,\mathrm{V}$ and $V_a=0.5\,\mathrm{V}$ for different
contacts in two cases: $t_2=1.1\,t_1$ (solid curve) and $t_2=t_1$
(dashed curve).}
\end{figure}

When the molecule is brought close to an electrode, the bonding
between them will depend on the molecule orientation. This
orientation can be such that only one carbon atom, a pentagon or a
hexagon, of the C$_{60}$ molecule be in contact with the leads.
Therefore, one can expect different conduction through the
molecule, which arises due to the interference effects, and will
be discussed in detail in the next section. For this reason, we
make use of the nonequilibrium Green's function technique to
obtain the current for a constant bias voltage $V_a$ between two
electrodes. Since the total Hamiltonian [Eq. (\ref{H})] does not
contain inelastic scatterings, the current is computed from the
Landauer formula \cite{Landa, Meir}
\begin{equation}\label{I}
I=\frac{2e}{h}\int_{-\infty}^{\infty}
T(\epsilon,V_a)[f(\epsilon-eV_a/2)-f(\epsilon+eV_a/2)]d\epsilon \
,
\end{equation}
where $f$ is the Fermi function and $T(\epsilon,V_a)$ is the
energy- and voltage-dependent transmission function given by
\begin{equation}\label{T}
T(\epsilon,V_a)=\mathrm{Tr}[\hat{\Gamma}_L(\epsilon-eV_a/2)\hat{G}_C(\epsilon,V_a)
\hat{\Gamma}_R(\epsilon+eV_a/2)\hat{G}_C^{\dagger}(\epsilon,V_a)]\
.
\end{equation}

The coupling matrices $\hat{\Gamma}_{L,R}$, also known as the
broadening functions, are related to the self-energies through
\begin{equation}\label{Gama}
\hat{\Gamma}_{L,R}=i[\hat{\Sigma}_{L,R}-\hat{\Sigma}^{\dagger}_{L,R}]\
.
\end{equation}

Equations (\ref{I}) and (\ref{T}) form the basis for our analysis
of the coherent transport through the molecular bridge. Our
approach, as a real-space method, makes it possible to model
arbitrary the number of contacts. In this regard, the core of the
problem lies in the calculation of the self-energies
$\hat{\Sigma}_{L,R}$. In the case of contact through a single
carbon atom of the molecule, only one element of the self-energy
matrices is non-zero. However, for the transport through opposite
pentagons or hexagons, 25 or 36 elements of the self-energy
matrices are nonzero, respectively.

\begin{figure}
\centerline{\includegraphics[width=0.8\linewidth]{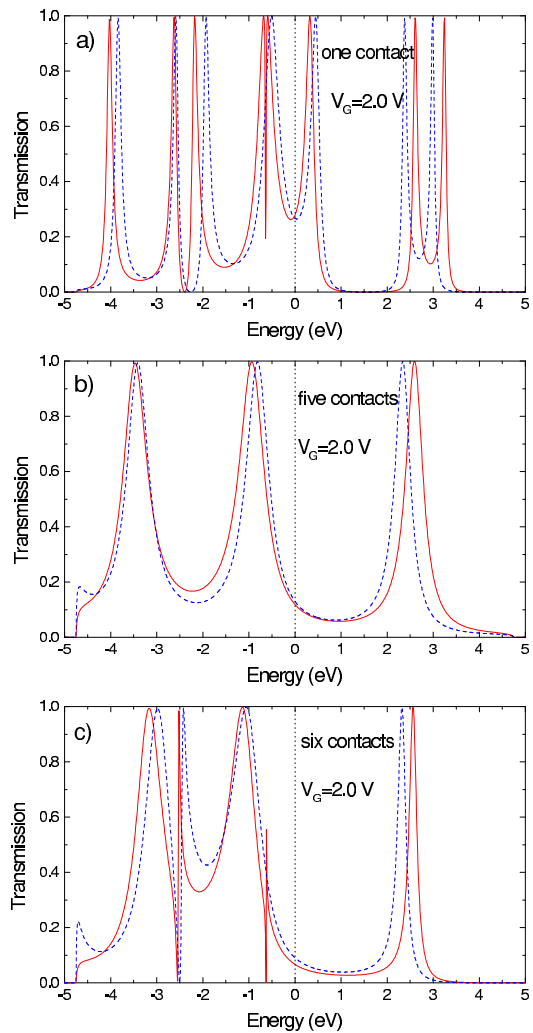}}
\caption{The same as Fig. 2 but for $V_G=2.0\,\mathrm{V}$.}
\end{figure}

\section{Results and discussion}
The present formalism can be applied to the systems in which an
arbitrary number of carbon atoms of the C$_{60}$ molecule can be
coupled to the one-dimensional metallic electrodes. In this study,
the coupling through one, five, and six carbon atoms to the
electrodes will be considered (see Fig. 1). These cases were
chosen as the most probable experimental orientations
\cite{Hashizume}. In the numerical calculations we set
$t_1=2.5\,$eV \cite{Paul}, $t_2=1.1\,t_1$ \cite{Manou},
$E_F=0.0\,$eV, and $T=300\,$K.

Connecting the molecule to the electrodes broadens and shifts the
discrete states of the molecule. This broadening of the states
depends on the strength of the coupling to the contacts as well as
the wave function of the particular state. Therefore, one can
expect different behaviors for the transmission coefficient (TC)
of the system in the cases of single and multiple contacts. In
order to investigate such a behavior, in Figs. 2 and 3 we have
shown the TC as a function of the energy of the electron which is
emitted from the left electrode into the molecule, for
$V_a=0.5\,$V at $V_G=0.0\,$V and $V_G=2.0\,$V, respectively. It is
clear that the transmission functions have large values (peaks)
near the molecular levels of C$_{60}$. In the coherent transport,
the electron wave function is assumed to extend coherently across
the whole system. Accordingly, when the electron energy nearly
coincides with the molecular levels, the electron resonantly
transmits through the molecule and a large transmission arises.
From Fig. 3, it is evident that, a gate voltage shifts the states
and, due to this effect, the transmission channels can
significantly vary . Therefore, a gate potential can change the
electron conduction through this device, producing field-induce
molecular switching.

The number of peaks in the case of connection to five carbon atoms
of the C$_{60}$ molecule is lower with respect to the other cases.
In the case of connection to one carbon atom, the value of TC at
the Fermi energy is near zero, however, it is considerable in the
case of five contacts. In both figures, the main factor for the
difference between the single and multiple contacts is the
interference effect. In fact, when the molecule is contacted
through one carbon atom to the electrodes, the transmission
through the molecule corresponds to the resonant tunneling
effects. With increasing the number of contact points, the
interference effects around these contact points become important,
some resonances might completely disappear, and the transmission
curve changes. By comparing the Figs. 2(b) and 2(c) with Fig.
2(a), and Fig. 3(b) and 3(c) with Fig. 3(a), one can easily
observe the influence of interference effects on the electronic
transport. The physical meaning of the interference effect is that
the electron waves in the molecule which come from the different
contact points may suffer a phase shift. Thus, a constructive or
destructive interference in the propagation process of the
electron through the molecule can occur. As mentioned in Sec. II,
the effect of contacts is described by the self-energy matrices.
Therefore, the Green's function (and hence the density of states)
of the coupled molecule and the transmission spectrum vary with
the number of contact points \cite{Datta}.

\begin{figure}
\centerline{\includegraphics[width=0.8\linewidth]{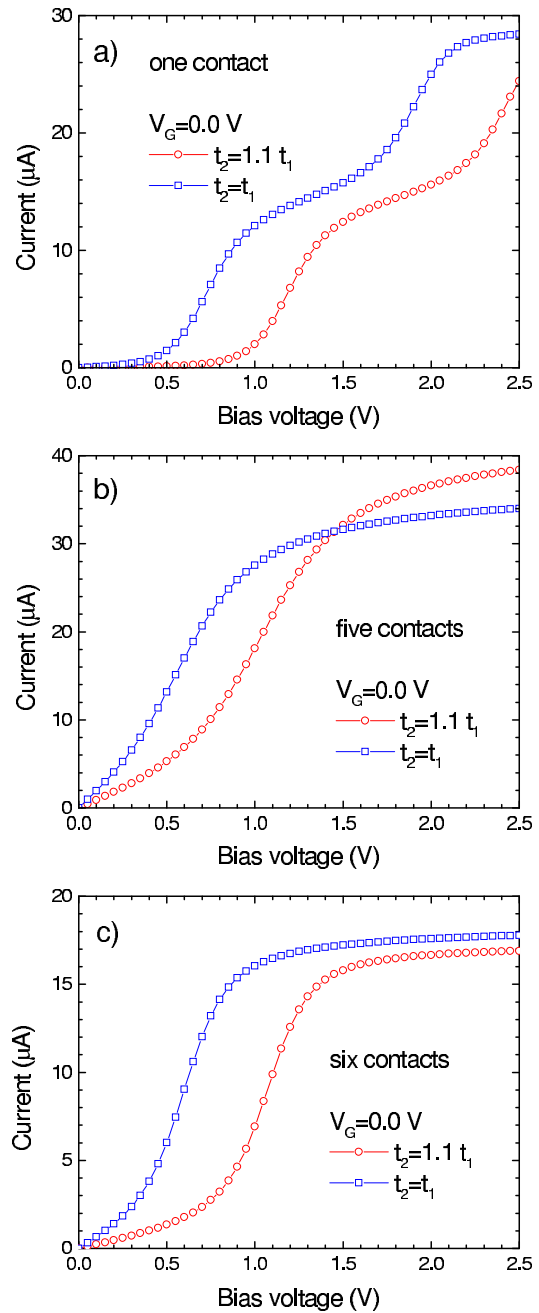}}
\caption{Current vs voltage characteristics at
$V_G=0.0\,\mathrm{V}$ for different contacts in two cases:
$t_2=1.1\,t_1$ (circle) and $t_2=t_1$ (square).}
\end{figure}

Note that in the quantum transport, the width of the transmission
resonances depends on the electrode/molecule coupling strength. In
one-dimensional electrodes, only one atom (end site) is coupled to
the carbon atoms. Therefore, if we increase the number of atoms at
the surface of electrodes, by choosing electrodes with finite
cross sections, then the width of resonances becomes significantly
broader for the cases of five and six atomic contacts
\cite{Palacios1}. We have also studied the effects of changing the
electrode/molecule coupling strength. The results for the case of
single contact showed that the TC increases with increasing $t'$.
For strong coupling, the peaks are broadened which indicate that
the electronic transport can no longer be considered as resonance
tunneling through eigenstates of the isolated C$_{60}$ molecule
\cite{Paul}. In the case of multiple contacts, however, the
broadening of peaks and the increase in TC are not noticeable.
This suggests that the interference effects play a dominant role
in the transmission through the molecule, when we increase the
number of contact points.

On the other hand, it is important to note that the presence of
both single and double bonds causes a shift in the peaks of TC.
Shift to the higher or lower energies depends on the peak
positions. From the Figs. 2(a) and 2(c), it is clear that the
difference in bond lengths gives rise to an extra peak in the TC
curves due to the appearance of a sharp dip at the energy near
-2.63 eV. The gate voltage shifts the position of peak, as shown
in Figs. 3(a) and 3(c). Therefore, the effect of bond dimerization
(i.e., $t_2$=1.1\,$t_1$) is that, the degeneracy of one of the
molecular levels is lifted \cite{Manou}. By applying a suitable
gate voltage, this effect may be important, if this level be
nearly coincided with the Fermi energy of the metallic electrodes.

\begin{figure}
\centerline{\includegraphics[width=0.8\linewidth]{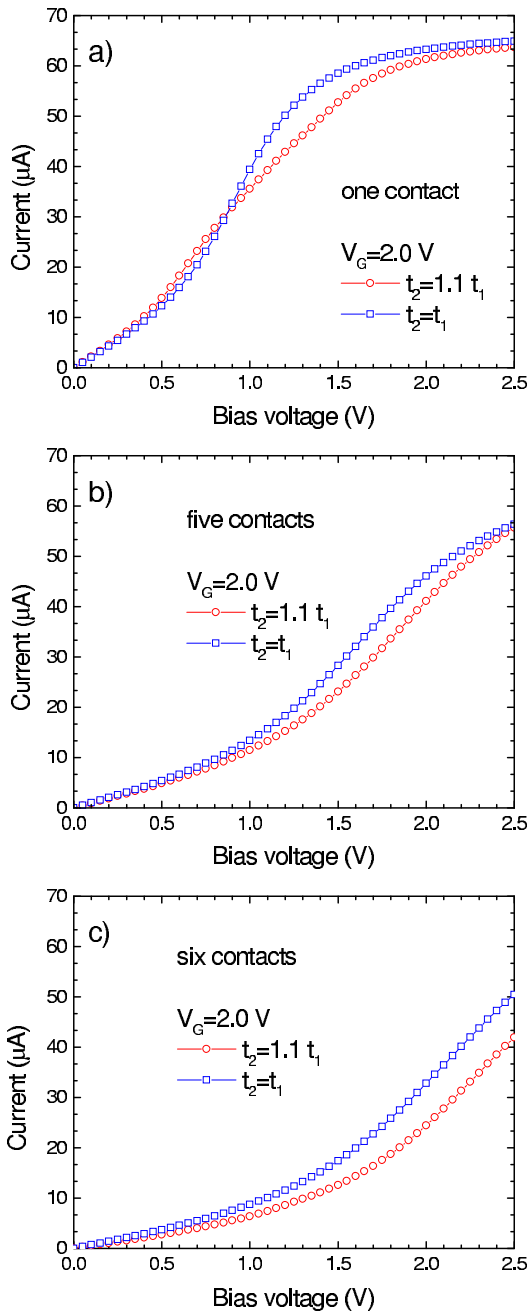}}
\caption{The same as Fig. 4 but for $V_G=2.0\,\mathrm{V}$.}
\end{figure}

In order to provide a deep understanding of the electronic
transport, we have shown the $I$-$V$ characteristics in Figs. 4
and 5, at $V_G=0.0\,$V and $V_G=2.0\,$V, respectively. Since our
structure is symmetric, we obtained a symmetrical behavior in the
$I$-$V$ curves with respect to $V_a=0.0\,$V. For this reason, we
have not shown the results for negative applied voltages. At
$V_G=0.0\,$V and in the case of one contact point, a threshold
voltage is needed to generate current through the device which
shows that, at low applied voltage the device is in its off state.
For a certain gate voltage such as $V_G=2.0\,$V, the device is
turned on and the current linearly increases for low applied
voltages [see Fig. 5(a)]. For the case that $t_2=1.1\,t_1$, the
threshold voltage is nearly two times larger than that in the case
of $t_2=t_1$. When the number of carbon atoms which take part in
coupling between the molecule and the electrodes increases, the
$I$-$V$ characteristics show an Ohmic behavior at low applied
voltages. Such a behavior is reasonable because in this situation,
the hybridization with the electrodes is stronger and there are
more paths for the electrons to pass from the metal to the
molecule \cite{Palacios1}. The curves in all figures, particularly
in Fig. 4, show a steplike behavior which indicates that a new
channel is opened.

\begin{figure}
\centerline{\includegraphics[width=0.8\linewidth]{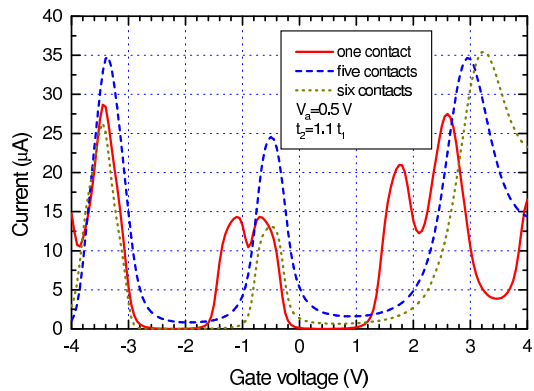}}
\caption{Current as a function of gate voltage at different
contacts in the case of $t_2=1.1\,t_1$.}
\end{figure}

The present results for the $I$-$V$ characteristics of the cases
of single and multiple contacts are qualitatively in agreement
with the experimental results in Refs.\,\cite{Por1,Por2}, and
\cite{Joach2} respectively. In Ref.\,\cite{Naka}, the bond length
difference has been ignored because they believed that the shift
in the energy spectrum is less than 2$\%$, and there is no
significant difference in the electron conduction through the
molecule. The present results, however, show that by adjusting the
parameters, a considerable difference in the $I$-$V$
characteristics between the cases of $t_2$=$t_1$ and
$t_2$=1.1\,$t_1$ may occur.

In Fig. 6, we have shown the current as a function of gate
voltage. The results reveal that the electron conduction can be
strongly dependent on the gate voltage. For the values of the gate
voltages that the current becomes zero, the molecule behaves as a
semiconductor and for the other values it acts as a metal. Such a
behavior can be clearly seen in the cases that one or five carbon
atoms of the molecule are contacted to the electrodes. Therefore,
the results suggest that the C$_{60}$ molecule is an interesting
candidate for operation of devices as a nanoscale current switch.

\section{Conclusions}
Using the nonequilibrium Green's function technique and the
Landauer theory, we have investigated the effects of different
contact geometries on the electronic transport through a C$_{60}$
molecule sandwiched between two metallic electrodes. It has been
shown that the transmission curves and the $I$-$V$ characteristics
calculated at the single and multiple couplings to the electrodes
can be completely different due to the interference effects. In
addition, the influence of gate voltage on C$_{60}$ and the effect
of different bond lengths in this molecule were observed as a
shift in the transmission peaks. Our results therefore indicate
that the contact geometries, bond dimerization, and
charge-transfer doping with gate voltage play important roles and
may change the physical picture of electron conduction in
C$_{60}$-based molecular devices.

\end{document}